\documentstyle[12pt]{article}
\begin{document}

\begin{center}
{\large{\bf Chiral nonperturbative approach to the isoscalar\\

\vspace{0.3cm}

s-wave $\pi \pi$ interaction in a nuclear  medium}}
\end{center}

\vspace{1cm}

\begin{center}
H.C. Chiang
\footnote{Permanent address: Institute of High Energy Physics. Chinese
Academy of Sciences, Beijing, 100039, China}, E. Oset and M.J. Vicente-Vacas
\end{center}

\vspace{0.1cm}

\begin{center}
{\small{\it Departamento de F\'{\i}sica Te\'orica and IFIC\\
Centro Mixto Universidad de Valencia-CSIC\\
46100 Burjassot (Valencia), Spain}}
\end{center}

\vspace{3cm}

\begin{abstract}
{\small{The  s-wave isoscalar amplitude for $\pi \pi$ scattering in a
nuclear medium is evaluated using a nonperturbative unitary 
 coupled channels  method and the standard chiral Lagrangians. The 
method has proved successful to describe the $\pi \pi$ properties
in the scalar isoscalar channel up to 1.2 GeV giving rise to poles
in the $t$ matrix for the $f_0 (980)$ and the $\sigma$. The
extension of the method to the nuclear medium implies not only the 
renormalization of the pions in the medium, but also the
introduction of interaction terms related to contact terms in the $\pi N 
\rightarrow \pi \pi N$ interaction.
Off shell effects are also shown to be important leading to 
cancellations which reduce the coupled channel integral equations to
a set of algebraic equations. As the density increases we find a reduction
of strength below the $\sigma$ region and a certain accumulation of
strength at energies around pion threshold. Our results, based
on  chiral Lagrangians, provide similar results to those
obtained with phenomenological models which impose minimal chiral 
constraints.}}
\end{abstract}

\newpage

\section{Introduction.}

The $\pi \pi$ interaction in a nuclear medium in the $J = T = 0$
channel ($\sigma$ channel) has stimulated much theoretical work lately.
It was realized that the attractive p-wave interaction of the pions
with the nucleus led to a shift of strength of the $\pi \pi$ system
to low energies and eventually produced a bound state of the two
pions around $2 m_\pi  -  10 \, MeV$ \cite{1}. This state would
behave like a $\pi \pi$
Cooper pair in the medium, with repercussions in several observable
magnitudes in nuclear reactions \cite{1}. The possibility that such
effects could have already been observed in some unexpected enhancement
in the ($\pi, 2 \pi$) reaction in nuclei \cite{2} was also noticed there.
More recent experiments where the enhancement is seen in the $\pi^+ \pi^-$
channel but not in the $\pi^+ \pi^+$ channel \cite{3} have added more
attraction to that conjecture.

Yet, it was early realized that constraints  of chiral symmetry might
affect those conclusions \cite{1}. In order to investigate the
influence of chiral constrain\-ts in $\pi \pi$ scattering in the nuclear
medium two different models for the $\pi \pi$ interaction were used
in \cite{4}.
One of them from \cite{5} did not satisfy the chiral constraints, while
another one from \cite{6} produced an amplitude behaving
like $m_\pi$ in the limit of small pion masses. The conclusion of
\cite{4} was that, although in the chirally constrained model
the building up of $\pi \pi$ strength at low energies was attenuated, it
was still important within  the approximations done in their calculations.
The latter ones used some approximations, amongst others, the use of only
$\Delta h$ excitation with zero $\Delta$ width to build up the $\pi$
nuclear interaction. Warnings
were also given that results might depend on the off shell
extrapolation of the $\pi \pi$ scattering matrix.

Further refinements were done in \cite{7}, where the width of the
$\Delta$ and coupling to $1p \, 1h$ and $2p \, 2h$ components were
considered. The coupling of pions to the $ph$ continuum led to a dramatic
re-shaping of the $\pi  \pi$ strength distribution, but the
qualitative conclusions about the accumulated strength at 
low energies remained.

In ref. \cite{8} the importance of the coupling to the $ph$ components
was reconfirmed and the use of more accurate models for the $\pi \pi$ 
interaction, as the J\"{u}lich model based on meson exchange \cite{9},
did not change the conclusions on the enhanced $\pi \pi$ strength at low
energies. However, the use of a linear and nonlinear models for the 
$\pi \pi$ interaction, satisfying the chiral constraints at small energies,
led to quite different conclusions and showed practically no
enhancement of the $\pi \pi$ strength at low energies. The same
conclusions were reached using the J\"{u}lich model with a
subtracted dispersion relation so as to satisfy the chiral constraints.
The latter model employed the Blakenbecler-Sugar equation in which the
$2 \pi$ intermediate states were placed on-shell. The conclusion
of this paper was that the imposition of chiral constraints in
the $\pi \pi$ amplitude prevented the pairing 
instabilities shown by the other models not satisfying those
constraints.

In a further paper \cite{10} the authors showed, however, that the
imposition of the chiral  constraints by themselves did not
prevent the pairing instabilities and uncertainties remained related to
the off-shell extrapolation of the $\pi \pi$ amplitude and the
possible ways to implement the minimal chiral constraints. The situation,
as noted in \cite{10} is rather ambiguous, but the studies done
have certainly put the finger in the questions that should
be properly addressed: chiral symmetry, off-shell extrapolations,
unitarity, etc.

After the above discussions it looks quite intuitive to
think that the use of chiral Lagrangians, and chiral perturbation
theory, where the $\pi \pi$ interaction at low energies has been
thoroughly studied \cite{11}, should be the appropriate framework
to look at this problem. However, the
scalar sector  shows additional problems. The
$J = T = 0 \; f_0 (980)$ resonance  does not show up in chiral perturbation
theory, and even the $\sigma$ pole does not show up in this
perturbative approach. Since one is dealing here with the
shift of the "$\sigma$  strength" to lower energies it looks most
advisable to start with a theory  where the $\sigma$ shows up neatly in 
the $\pi \pi$ interaction. Fortunately, two recent independent
approaches using the Gasser and Leutwyler chiral Lagrangians,
which implement unitarity in an exact way, have succeeded in
reproducing the low energy $\pi \pi$ phase shifts while at the same
time generating a "$\sigma$" pole in the $\pi \pi$  $T$ matrix in the
II sheet of the complex plane \cite{12,13}.

In ref. \cite{12}  the method of the inverse amplitude of  \cite{14}
was used where elastic unitarity in the $\pi \pi$ channel is imposed.
The method obtains good results for the low energy $\pi \pi$ interaction
in all channels. It, however, fails to obtain the $f_0 (980)$ and $a_0 (980)$
resonances in the scalar $T = 0$ and $T = 1$ channels
respectively, but the $\sigma$ pole in $J = T = 0$ is obtained.

In ref. \cite{13} unitarity in coupled channels is built from the
beginning with $\pi \pi$ and $K \bar{K}$ in the scalar, isoscalar
sector and $\pi \eta, K \bar{K}$ for $J = 0, T = 1$.  The phase
shifts and inelasticities are well reproduced up to about 1.2 GeV. The
$\sigma$ and $f_0 (980)$ resonances appear as poles in the $J = T = 0$
channel and the $a_0 (980)$ appears as a pole in $J = 0, T = 1$.
The coupling of channels was found essential to produce the $f_0 (980)$
and $a_0 (980)$ resonances, while the $\sigma$ pole was not much
affected by the coupling of the pions to $K \bar{K}$. This would
make the approaches of \cite{12} and \cite{13} similar at energies around
the $\sigma$ pole and, indeed,
 the results in that region are practically equal.
This has been made more explicit in a recent paper where the two
methods discussed above are unified into a more general scheme
\cite{14a}. A different perspective of this method is also given in
\cite{14b}.

The existence of these chiral nonperturbative methods offers unique
opportunities to tackle the problem of the scalar isoscalar $\pi \pi$
interaction in the nuclear medium and this is the purpose of the present
work. We follow here the approach of ref. \cite{13}, where one of the
problems pointed above, the off shell extrapolation, was worked out 
with detail. This, together with the automatic implementation of chiral
symmetry and its breaking, given by the Gasser and Leutwyler chiral
Lagrangians, allows us to face the problems mentioned above from a 
different perspective, where chiral symmetry, off shell extrapolation, 
etc., are directly associated to the structure of the chiral Lagrangians.

In constructing the $\pi \pi$ amplitude in the medium we will show
that chiral symmetry introduces new terms which do not appear in the
approaches discussed above and which lead to cancellations
of other terms coming from the off shell extrapolation of the $\pi \pi$ 
amplitude.

The results show some enhancement of the $\pi \pi$ strength at low
energies, similar to what is found in different approaches, particularly
those imposing the minimal chiral constraints. The work serves
to establish closer links with chiral dynamics and justify certain
prescriptions proposed in the past.

\section{Non perturbative chiral approach to $\pi \pi$ scattering
in the isoscalar isovector channel.}

We briefly summarize here the ingredients of ref. \cite{13} which will
be used here. The approach uses the coupled channels Lippmann 
Schwinger (LS) equation, although with relativistic meson 
propagators (equivalent to Bethe Salpeter equations). 
We take the states $|1 > = (K \bar{K}, T = 0>, 
| 2> = | \pi \pi , T = 0>$ and the LS equations read as

\begin{equation}
T_{ij} = V_{ij} + \overline{V_{il} G_{ll} T_{lj}}
\end{equation}

\noindent
where $V_{ij}$, the potential or Kernel of the LS equations, is
obtained from the lowest order chiral Lagrangians \cite{11}.

\begin{equation}
L_2 = \frac{1}{12 f^2} < (\partial_\mu \Phi \Phi - \Phi \partial_\mu
\Phi)^2 + M  \Phi^4 \, >
\end{equation}

\noindent
where the symbol $< >$ indicates the trace in flavour space of the $SU (3)$ 
matrices,  $f$ is the pion decay constant and $\Phi$, $M$ are $SU (3)$ 
matrices given by 

\begin{equation}
\begin{array}{l}
\Phi \equiv \frac{\vec{\lambda}}{\sqrt{2}} \vec{\phi} = \left(
\begin{array}{ccc}
\frac{1}{\sqrt{2}} \pi^0 + \frac{1}{\sqrt{6}} \eta_8 &
\pi^+ & K^+\\[2ex]
\pi^- & - \frac{1}{\sqrt{2}} \pi^0 + \frac{1}{\sqrt{6}} \eta_8 & K^0\\[2ex]
K^- & \bar{K}^0 & - \frac{2}{\sqrt{6}} \eta_8
\end{array} \right) \quad , \\[4ex]
\\
M = \left( \begin{array}{ccc}
m_\pi^2 & 0 & 0\\
0 & m_\pi^2 & 0\\
0 & 0 & 2m_K^2 - m_\pi^2 \end{array} \right) \quad .
\end{array}
\end{equation}
In the mass matrix, $M$, we have taken the isospin limit $(m_u = m_d)$.

The elements of $V_{ij}$ in the s-wave and $T = 0$ needed here are given
in \cite{13} by

\begin{equation}
\begin{array}{l}
T = 0\\
V_{11} = - < K \bar{K} | {\cal L}_2 | K \bar{K} > = - \frac{1}{4 f^2}
(3 s + 4 m_K^2 - \sum_i p_i^2)\\[2ex]
V_{21} = - < \pi \pi | {\cal L}_2 | K \bar{K} > = - \frac{1}{3 \sqrt{12} f^2}
( \frac{9}{2}s + 3 m_K^2 + 3m_\pi^2 - \frac{3}{2}  \sum_i p_i^2)\\[2ex]
V_{22} = - < \pi \pi | {\cal L}_2 | \pi \pi  > = - \frac{1}{9  f^2}
(9 s + \frac{15  m_\pi^2}{2}  - 3 \sum_i p_i^2)
\end{array}
\end{equation}

In eq. (1)   the term $\overline{VGT}$ stands for the integral

\begin{equation}
\overline{VGT} = \int \frac{d^4 q}{(2 \pi)^4} 
V (k_1, k_2, q) G (P,q) T (q; k'_1, k'_2)
\end{equation}

\noindent
where $k_1,k_2 (k'_1, k'_2)$ are the initial (final) momenta of the mesons,
$P = k_1 + k_2 = k'_1 + k'_2$ is the total momentum of the meson-meson system
and $q$ is the loop variable in the diagrams implicit in eq. (1) which
are depicted in fig. 1.

$G (P,q)$ is now the product of the two meson propagators

\begin{equation}
G_{ii} = i \frac{1}{q^2 - m^2_{1 i} + i \epsilon} \quad 
\frac{1}{(P - q)^2 - m^2_{2i} + i \epsilon} 
\end{equation}

The on shell values of $V_{ij}$ from eq. (4) are obtained
substituting $p^2_i = m_i^2$. The off
shell extrapolation is then given by

\begin{equation}
V_{off} = V_{on} + \beta \sum_i (p_i^2 - m_i^2)
\end{equation}

This peculiar off shell dependence has a practical consequence which
converts the integral LS equations into ordinary algebraic equations.
This is discussed in detail in ref. \cite{13} but can be envisaged
here quickly in the following way. Take a one loop diagram in
the series of fig. 1 which will involve 
$V_{off}^2$. This latter quantity can be written as 

\begin{equation}
V^2 = V^2_{on} + 2 \beta V_{on} \sum_i (p^2_i - m_i^2) + \beta^2
\sum_{ij} (p^2_i - m_i^2) (p^2_j - m^2_j) \,.
\end{equation}

Take the second term of the right hand side of eq. (8). The term
$p^2 - m^2$ just kills one of the propagators in $G$ of eq. (6).
The remaining quantity can be easily integrated and gives a term 
of the type 
$V_{on} q_{max}^2$, where $q_{max}$ is the cut off in the three
momentum. The interesting thing to observe is that this term
has the structure of the tree level term in the series, $V_{on}$, and
hence is incorporated in the potential by means of a renormalization
of the coupling $f$.

The use of the physical value of $f$ incorporates effectively that term 
which,
hence, must not be included in the calculation. Similarly the third term in
eq. (8) can be reabsorbed in the coupling and the masses of the
particles \cite{13}. The practical consequence of this is that only
the on-shell part of $V_{ij}$ must be kept in the loop integral and,
since they do not depend on $q$, they factorize outside the 
integral. The procedure can be  repeated to higher order
loops and thus the coupled LS equations become ordinary algebraic
equations given by

\begin{equation}
T_{ik} = V_{ik} + V_{ij} G_{jj} T_{jk}\\[2ex]
\end{equation}

\noindent
where

$$
G_{jj} = i \int \frac{d^4 q}{(2 \pi)^4} \frac{1}{q^2 - m_{1j}^2 + i \epsilon}
\quad \frac{1}{(P - q)^2 - m_{2j}^2 + iG}
$$

\noindent
and hence, in matrix form

\begin{equation}
T = [1 - VG]^{-1} V
\end{equation}

\section{ $\pi \pi$ scattering in the nuclear medium.}

We will follow the previous approaches and will renormalize  the pion
propagators in $G$. In our coupled channel approach we should also 
renormalize the kaons, but given the fact found in \cite{13} that
the $K \bar{K}$ system does not affect much the low energy $\pi \pi$
regime, we shall keep the $K \bar{K}$ state in our coupled channel
approach but without renormalization.

The pions are renormalized by allowing them to excite $ph$ and $\Delta
h$ (the $\Delta$ with a finite width). At the level of one
loop our renormalized amplitude would now contain the diagrams of fig. 2. 
Now let us take for instance the diagram of fig. 2b, cut it by a vertical
line that cuts simultaneously the $ph$ and the lower pion line, and
keep the part of the diagram 
to the left of this vertical line. We obtain the diagram 
of fig. 3a. This diagram can be interpreted
as one term contributing to the $\pi N \rightarrow \pi \pi N$ amplitude.
However, even before chiral perturbation theory established an elegant
and practical way to implement chiral symmetry and its breaking, it 
was already known that chiral symmetry required an extra term, shown in
fig. 3b, which was readily obtained from a set of chiral Lagrangians 
\cite{15}. Actually, one of the powerful consequences of chiral
symmetry is that it establishes a relationship between amplitudes
with different number of meson as external particles.

The set of the chiral Lagrangians is not unique and unitary transformations
or redefinition of fields are possible \cite{16}. The interesting
thing is that, while each one of the terms in fig. 3 depends on the
choice of Lagrangians, the sum of the two is independent of it.
This is the case of the $\pi N \rightarrow \pi \pi N$ amplitude as
well  as any related observable magnitude. An example of it is
found in the evaluation of the contribution  of the nuclear virtual pion cloud
to the pion selfenergy in the nuclear medium \cite{17}.

Chiral perturbation theory has allowed us to obtain the pion pole term
of fig. 3a and the contact term of fig. 3b in an easier way than done
in the past and also has allowed the possibility to extend these
ideas to the octet of pseudoscalar mesons. As an example, the pion pole
term and contact term for the $K \bar{K} \pi NN$  vertex were evaluated
in \cite{18}, where the contribution of the virtual pion cloud to $K^+$
nucleus scattering was investigated.

The contribution of the diagrams of fig. 3a, 3b is the starting point
of all models for the $\pi N \rightarrow \pi \pi N$ reaction
\cite{16,19,19a,20,21} as well as in the $K N \rightarrow K \pi N$ reaction
\cite{22}.

 The requirements of chiral symmetry force us to include the contact
term together with the pion pole term, and some cancellations appear that 
make the physical amplitudes respect the chiral limits, even in the 
presence of the nuclear medium. Indeed, in ref. \cite{18} an exact 
cancellation was found  between the contributions related to the
pion pole and contact terms, while in ref. \cite{17} the contribution
was finite but vanished in the limit of $m_\pi \rightarrow 0$.

The former discussion has shown us that in addition to the terms 
depicted in fig. 2 we must add the terms depicted in fig. 4.

The Lagrangians involving the contact terms are obtained from the
general chiral Lagrangians involving the pseudoscalar meson and baryon
octets \cite{11,23,24,25}

\begin{equation}
\begin{array}{rl}
{\cal L}_1^{(B)} & = < \bar{B} i \gamma^\mu \nabla_\mu B > - M_B < \bar{B}
B>\\[2ex]
 + & \frac{D + F}{2} < \bar{B} \gamma^\mu \gamma_5 u_\mu B > + 
\frac{D - F}{2} < \bar{B} \gamma^\mu \gamma_5 B u_\mu \,> ,
\end{array}
\end{equation}

\noindent
where $B$ is a $3 \times 3$ matrix, which in our case,
where only protons and neutrons are involved, reads as

\begin{equation}
B = \left( \begin{array}{ccc}
0 & 0 &  p\\
0 & 0 & n\\
0 & 0 & 0 \end{array}  \right)
\end{equation}

The three pion and a nucleon vertices, fig. 3b, are derived in \cite{18}
with the result

\begin{equation}
\begin{array}{ll}
{\cal L}_1^{(B)} & = \frac{D + F}{2} (\bar{p} \gamma^\mu \gamma_5 u_\mu^{11}
p + \bar{n} \gamma^\mu \gamma_5 u_\mu^{22} n\\[2ex]
& + \bar{n} \gamma^\mu \gamma_5 u_\mu^{21} p + \bar{p} \gamma^\mu
\gamma_5 u_\mu^{12} n)\\[2ex]
& + \frac{D - F}{2} (\bar{p} \gamma^\mu \gamma_5 u^{33}_\mu +
\bar{n} \gamma^\mu \gamma_5 u_\mu^{33} n)\, ,
\end{array}
\end{equation}

\noindent
where $u^{ij}_\mu$ denotes the $(i,j)$ matrix element of the $u_\mu$
matrix defined as

\begin{equation}
\begin{array}{ll}
u_\mu & = - \frac{\sqrt{2}}{f} \partial_\mu \Phi\\[2ex]
& + \frac{\sqrt{2}}{12f^3} (\partial_\mu \Phi \Phi^2 - 2 \Phi \partial_\mu
\Phi \Phi + \Phi^2 \partial_\mu \Phi)\\[2ex]
 & + O (\Phi^5) \, .
\end{array}
\end{equation}

The $(D - F)$ term in eq. (13) does not contribute in our case since
$u_\mu^{33}$ contains kaon fields.

By using the nonrelativistic reduction $\gamma^\mu \gamma_5
p_\mu \rightarrow - \vec{\sigma} \vec{p}$,  the relevant terms which are
needed in our approach are  evaluated and they are shown in the Appendix.

We can also generalize these vertices to the case of $N \Delta$ transition
by substituting \cite{18}

\begin{equation}
\sigma_i \tau_j (\hbox{for nucleons}) \rightarrow
\frac{f^* _{\pi N \Delta}}{f_{\pi N N}} S_i^\dagger T^\dagger_j
\end{equation}

\noindent
where $S^\dagger, T^\dagger$ are the spin, isospin transition operators
from 1/2 to 3/2.

As an example let us write the contribution of the left hand side
vertex of the diagram of fig. 4a, which we depict in fig. 5a with labels
for the momenta and a particular choice of pion charges. In the 
$\pi^+ \pi^-$ CM frame $(\vec{k}_1 + \vec{k}_2 = 0)$, and projecting
over s-wave, we obtain the contribution of this vertex

\begin{equation}
- i \tilde{t}_{ph} = i \frac{1}{6 f^4} (\frac{D + F}{2})^2
\vec{q}_2\,^2 U_N (q_2)
\end{equation}

\noindent
where $U_N$ is the Lindhard function for $ph$ excitation \cite{26}.
The use of the Lindhard function accounts for forward
and backward propagating bubbles and hence we are automatically
taking into account the two diagrams depicted in fig. 5, where the proton
of the $ph$ excitation is an occupied state in diagram (a) while the
neutron is the occupied state in diagram (b).
It is straightforward to take into account the $\Delta h$ excitation. It is
sufficient to substitute $U_N$ by $U_N + U_\Delta$, where $U_\Delta$ is
the Lindhard function for $\Delta h$ excitation conveniently normalized.
Formulae for $U_N, U_\Delta$ with the normalization required here can be
found in the Appendix of ref. \cite{27}.

The next step requires the evaluation of this vertex in the isospin
state $T = 0$.
The $T = 0$ state is

\begin{equation}
| \pi \pi, T = 0> = - \frac{1}{\sqrt{6}} | \pi^+ (\vec{q}) \pi^-
(- \vec{q}) + \pi^- (\vec{q}) \pi^+ (- \vec{q}) + \pi^0 (\vec{q})
\pi^0 ( - \vec{q}) >
\end{equation}

\noindent
where the phases and normalization are chosen as in \cite{13}.
The extra factor $\frac{1}{\sqrt{2}}$ in the normalization is chosen
such as to preserve the closure sum, $\sum_{\vec{q}} | > < | = 1$,
because,  the $|\pi \pi, T = 0> $ is a symmetrical state.

By summing the contributions from the $ph$ and $\Delta h$ on the upper 
 meson line we obtain

\begin{equation}
\tilde{t} = - \frac{1}{3 f^4} (\frac{D + F}{2})^2 \vec{q}\,^2 \,
U (q_2)
\end{equation}

\noindent
where $\vec{q} = \vec{q}_1 = - \vec{q}_2$.

Next we turn our attention to another sort of diagram which we obtain
from the consideration of the contact term in each one of the vertices
of the one loop diagram. This is depicted in fig. 6. Its contribution
to the $\pi^+ \pi^-$ $T$ matrix with a $\pi^+ \; \; ph$ intermediate
state
 is readily evaluated and one obtains,

\begin{equation}
\tilde{t}_{R, ph} = i \frac{1}{ 36 \, f^6} (\frac{D + F}{2})^2 \int
\frac{d^4 q_1}{(2 \pi)^4} \vec{q}_1\,^2 D_0 (q_1) U_N (q_2)
\end{equation}

\noindent
where $D_0 (q_1)$ is the pion propagator and $q_2 = k_1 + k_2 - q_1$.

 Furthermore, substituting the
$\pi^+ \; ph$ intermediate state by
a $\pi^0 \; ph$ state leads to a similar contribution but substituting
$\frac{1}{36} $ by $\frac{1}{18}$ in eq. (19),
account taken of the symmetry of the intermediate state which introduces
a relative factor $\frac{1}{2}$ since the diagram where the $ph$
is excited in the lower one is topologically equivalent to the
former one.
 One can work out the
other combinations with $\pi^0 \pi^0$ in the initial or final
states and then evaluate the $T = 0$ contribution which is given by

\begin{equation}
\tilde{t}_R = i \frac{1}{9 f^6} (\frac{D + F}{2})^2 \int
\frac{d^4 q_1}{(2 \pi)^4} \, D_0 (q_1) \vec{q}_1\,^2 \, U_N (q_2)
\end{equation}

Once again, taking into account $\Delta h$ excitation is straightforward
and one simply substitutes $U_N$ by $U_N + U_\Delta$ in eq. (20).

\section{Off shell extrapolation of amplitudes and cancellations.}

Let us come back to the diagram of fig. 2b. In the case of  free pion
scattering we could prove that the $\pi \pi$ amplitude in the  loops 
factorized on-shell, and the off shell part went into renormalization of
couplings and masses\cite {13}. Here the presence of the ph excitation changes the
analytical structure of the diagram and we must investigate what happens
to the off shell extrapolation of the $\pi \pi$ amplitudes. For this purpose
we recall that we shall be interested in the strength of the $\pi \pi$
system, which is related to $Im  T_{22}$. If we look at the diagram
of fig. 2b, the imaginary part can come, according to Cutkosky rules, when
the two intermediate pions are placed on-shell or when the lower pion and the
ph are placed on-shell. In both cases the lower pion with momentum $q_1$
is placed on-shell . The same occurs with the
diagrams of fig. 4a,b. Thus, the off-shell $\pi \pi$ potential
in those diagrams, with external pions placed on-shell, according to
eqs. (4), (8), reads now

\begin{equation}
V_{off} = V_{on} + \frac{1}{3 f^2} (q^2_2 - m_{\pi}^2)
\end{equation}

The contribution of the diagram  of fig. 2b  to the $\pi \pi$  S = 0, T = 0
amplitude is given by

\begin{equation}
\begin{array}{ll}
- i t^{(1)} & = \int \frac{\displaystyle{d^4 q_1}}{\displaystyle{
(2 \pi)^4}} \, \frac{\displaystyle{1}}{\displaystyle{f^2}} \,
\{ V^2_{on} + \frac{2}{3 f^2} V_{on} (q^2_2 - m_{\pi}^2) +
\frac{1}{9 f^4} \, (q^2_2 - m_{\pi}^2)^2 \} \\[2ex]
& \times (\frac{D + F}{2})^2 \, D^2_0 (q_2) D_0 (q_1) \vec{q} \, ^{2}_2 
U (q_2)
\end{array}
\end{equation}

On the other hand, the sum of the amplitudes of diagrams (a) and (b) of
fig. 4,  which contribute equally, is given by

\begin{equation}
\begin{array}{ll}
- i t^{(2)} = & - 2 \int \frac{\displaystyle{d^4 q_1}}{\displaystyle{(
2 \pi)^4}} \, \frac{\displaystyle{1}}{\displaystyle{3 f^4}} \,
\left( \frac{\displaystyle{D + F}}{\displaystyle{2}} \right)^2 \, \vec{q} \,
 ^{2}_2 D_0 (q_1) D_0 (q_2) \,
   \{ V_{on} + \frac{1}{3 f^2} \, (q_2^2 - m^2_{\pi}) \}\\[2ex]
& \times U (q_2)
\end{array}
\end{equation}

It is interesting to note that the terms  proportional to $V_{on}$ in eq. 
(22)
and eq. (23) cancel exactly. This leaves us with the term with $V^2_{on}$
in eq. (22) plus the terms proportional to $(q^2_2 - m^2_{\pi})^2$ in eq. (22)
and the one proportional to $(q^2_2 - m^2_{\pi})$ in eq. (23). It is also
interesting to note that these latter two terms have the same structure as
the term $\tilde{t}_R$ from eq. (20), corresponding to the diagram of 
fig. 6, and the sum of the three also cancels exactly.

In practical terms the situation has become rather easy. The terms of fig. 4
and 6
do not have to be evaluated and those of fig. 2 must be included but with 
$V_{\pi \pi}$  evaluated on-shell. These findings agree with the results
of \cite{28a} which also show that the off shell contribution depends
on the representation chosen while observable quantities should be
independent of it.

The arguments can be extended to higher order loops of the type of fig. 2
and fig. 4, with the result that we must omit the terms of the type of fig. 4
and 6
and include only
loops of the type of fig. 2 but with $V_{\pi \pi}$ on-shell. This
allows the factorization of the potential outside the integrals and the
Lippmann Schwinger equations are readily evaluated
since they become algebraic equations like in the free
 $\pi \pi$  scattering case.

\section{Coupled channel equations}

We take as channels $\pi \pi$ and $K \bar{K}$ but do not renormalize the
$K \bar{K}$ system as discussed  above. The
 series of terms in the Lippmann Schwinger
eqns., which include  the potential, the terms of fig. 2 and higher order 
iterations
of that type, including also free $K \bar{K}$ intermediate states, is given by

\begin{equation}
T_{22} = V_{22} + V_{21} G_{11} t_{12} + V_{22} \tilde{G}_{22}
T_{22}
\end{equation}

\noindent
where only $T_{22}$ and $\tilde{G}_{22}$ are renormalized in the 
medium. The other quantities are evaluated in free space and 
are taken from ref.
\cite{13}. We can obtain $T_{22}$ from eq. (24) 

\begin{equation}
T_{22} = \frac{V_{22}+  V_{21} G_{11} t_{12}}{1 -  V_{22}\tilde{G}_{22}}
\end{equation}

The  integral of the two pion propagators in the medium, $\tilde{G}_{22}$,
 is then given by

\begin{equation}
\tilde{G}_{22} = i \int \frac{d^4 q}{(2 \pi)^4} D (q) D (P - q)
\end{equation}

\noindent
with

\begin{equation}
V_{22} = V_{22, on}  = - \frac{1}{9 f^2} (9 s + \frac{15 m_\pi^2}{2} -
12 m_\pi^2)
\end{equation}

\noindent
and

\begin{equation}
D (q) = \frac{1}{q^2 - m_\pi^2 - \Pi (q)}
\end{equation}

\noindent
where $\Pi (q)$ is the pion selfenergy in the medium

\begin{equation}
\Pi (q) = \frac{(\frac{D+F}{2 f})^2 \vec{q} \, ^{2}
U (q)  }{1 - (\frac{D+F}{2 f})^2 g' U (q) }
\end{equation}

\noindent
with $g'$ the Landau-Migdal parameter, which we take as $g' = 0.7$.

When including the pion selfenergy in eq. (28) we also go beyond the lowest
order diagrams in density that we have discussed in detail, but this is the
appropriate way to take the higher order diagrams into account.

We have also included the pion selfenergy accounting for $2 p 2h$
excitation.
Since we are concerned mostly around the pion threshold region
($\sqrt{s} \sim 2 m_\pi)$ we have taken this selfenergy from pionic
atoms. The procedure is discussed in detail in \cite{28b} and it
amounts to substituting in eq. (29)

\begin{equation}
\left( \frac{D + F}{2 f} \right)^2 U(q) \rightarrow \left(
\frac{D + F}{2 f}\right)^2 U (q)
- 4 \pi C_0^* \rho^2
\end{equation}

\noindent
with $\rho$ the nuclear density and

\begin{equation}
C_0^* = (0.105 + i   \, 0.096) \, m_\pi^{-6}
\end{equation}

It is interesting to note that eq. (25) leads to a zero very close
to the one where $V_{22} = 0$ (Adler zero) since the second term in the
numerator of eq. (25), involving kaon loops, is negligible around
that energy. Hence our approach
fulfils the minimal chiral constraints (MCC) of ref. \cite{10} even
in the presence of the nuclear medium.

\section{Results and discussion}

In fig. 7 we show $Im T_{22}$ as a function of the energy for
different values of $k_F$, the Fermi momentum. The results
show a reduction of strength below the `$\sigma$' region and an
accumulation of strength at low energies around the
pion threshold. The results obtained omitting the $2p 2h$ part of the
pion selfenergy are qualitatively similar to those in fig. 7 at energies
below 600 $MeV$. In the region close to the dip of the $f_0 (980)$
resonance the results including the $2p 2h$ part are about $30 \, \%$
smaller than those omitting them. At these energies a more realistic 
evaluation of this part would be needed but since we are not concerned
about this region we do not elaborate further on the issue.

The results obtained resemble very much those shown in fig. 12 of ref.
\cite{8} and particularly those of fig. 3 of \cite{28c}, where first
attempts to relate the accumulated  strength around pion threshold
to the experiment of \cite{3} are done.

 The association
of the peaks found here to the extra
strength around pion threshold found in the experiment at small pion pair 
invariant mass \cite{3} is not straightforward. Indeed,
even if this invariant mass is small, the pion pair moves  with some momentum.
Here we have studied the $\pi \pi$ system at rest in nuclear matter and there
could be differences for a $\pi \pi$ pair moving with respect to the rest
frame of the Fermi sea. Yet the steps taken in \cite{28c} are encouraging
and more work along these lines would be welcome.

It should also be pointed out that we have selected the diagrams for
$\pi \pi$ free scattering and carried out the renormalization for the
mesons while at the same time have kept the partner terms which
appear at the same order of the chiral counting and make the scheme
invariant under unitary transformations of the fields. This would in
principle guarantee that relationships like Ward identities and other,
which are sometimes used in the study of the interaction of $\rho$-mesons
with matter \cite{29}, would be satisfied. Alternative methods to test Ward 
identities can be made using the master formula of \cite{30}, where the
most general amplitude for $\pi \pi$ scattering satisfying chiral 
constraints is derived and expressed in terms of form factors and 
polarization functions.
The use of resonance saturation for the evaluation of these latter magnitudes
leads to a consistency check of the master formula \cite{30}. Our
approach \cite{13,14a} leads dynamically to the different meson meson
resonances and would naturally provide the resonance contribution  to
those functions, hence indirectly fulfilling the test done in \cite{30}.

On the other hand, one can think of other many body diagrams which in
principle contribute to the process. Think for instance of a baryon
box diagram with 4 meson legs attached at different points of it, or
the same box diagram with two pairs of mesons attached at two points,
etc. These diagrams require the use of other terms of the chiral
Lagrangians used above and do not interfere with the counting
done so far. Furthermore, inspection of these diagrams proves that sometimes
the isoscalar $\pi N$ amplitude is involved, which is quite small on shell,
although it can be appreciably modified in some off shell situations.
Other times one meets with  the $p$-wave $\pi N$ interaction, while 
we are interested in s-wave propagation. In most cases, like in those box 
diagrams, a $ph$ is forced to propagate carrying zero momentum and the two
meson energy, which places it very far off shell, etc. These are qualitative
arguments which indicate small contributions from such terms and
would support our choice of many body diagrams as the
relevant set for the process studied. However, detailed studies of these
alternative many body mechanisms would be welcome.

\section{Conclusions}

We have performed calculations of the $\pi \pi$ scalar-isoscalar amplitude
in a nuclear medium starting from the standard chiral Lagrangians and using
a unitary framework with coupled channels which proved rather successful
in describing the meson-meson interaction in the scalar sector.

Compared to other schemes that impose minimal constraints of chiral symmetry,
essentially the vanishing of the $\pi \pi$ amplitude in the limit of
$m_{\pi} \rightarrow 0$, our scheme uses the input of the standard chiral 
Lagrangians and generates different terms, in the expansion on the number
of meson fields, which appear on the same footing in the $\pi \pi$
amplitude in the presence of a nuclear medium. In this way, some terms related
to the contact term $\pi \pi \pi N N$, which appears in the
$\pi N \rightarrow \pi \pi N$ reaction, and which are new with
respect to previous approaches, are generated here. Simultaneously, the 
off-shell extrapolation
of the chiral $\pi \pi$ amplitudes is used and it is shown to produce 
cancellations with the terms coming from the contact vertices, with
the remarkable result that only the one shell part of the
$0 (p^2)$ meson meson amplitudes is needed, like in the case
of free meson scattering.

This shows the usefulness of working explicitly with chiral Lagrangians
in order to find out those subtle cancellations. Our results show
a reduction of strength of $Im T_{22}$ in the `$\sigma$' region
and an accumulation of strength close to pion threshold which
are features shared by most of the approaches.

Quantitatively our results resemble very much some results
in the literature which use models imposing minimal chiral constraints.
Among a large variety of prescriptions used in the past to account
for $\pi \pi$ interaction in the medium, the present approach offers
a cleaner link to chiral dynamics. This has been made possible by
the work of \cite{13} which could combine exact unitarity with the
input of the chiral Lagrangians, describing accurately the free meson
meson interaction. Work along these lines in the study of the
modification of the meson meson interaction in a nuclear medium
for other channels is equally possible and
would be welcome.

On the other hand, the clear accumulation of strength around pion
threshold (even without singularities as claimed in some approaches) which
is also shared by some models, is an appealing feature that most probably
can be linked to present experiments showing enhanced distributions
of $\pi \pi$ invariant mass in $T = 0$ around pion threshold. Further
investigation along these lines is another of the challenging tasks
ahead.

\vspace{1cm}

Acknowledgements:

\vspace{0.5cm}

We would like to thank J. A. Oller for help and discussions.
Discussions with Z. Aouissat, B. Friman, J. Wambach, A. Wirzba, G.
Chanfray and I. Zahed are also acknowledged.
 One of us, H. C.
Chiang wishes to acknowledge financial support from the Ministerio de 
Eduaci\'on y Cultura in his sabbatical 
stay in the University of Valencia. This work is partly supported
by DGICYT contract number PB96-0753.

\newpage

\appendix
\section*{Appendix}

Matrix elements of the contact terms of fig. 8 in the meson-meson  CM frame 
($\vec{k}_1+\vec{k}_2 = 0$):

\begin{equation} 
  i L_{a} =  {{D+F}\over{2}} {{\sqrt{2}}\over{12 f^3}}\, 2\;
 \vec{\sigma}\cdot ( 3\vec{k}_1+\vec{q}_1),
\end{equation}

\begin{equation} 
  i L_{b} =  {{D+F}\over{2}} {{\sqrt{2}}\over{12 f^3}} \, 2\;
 \vec{\sigma}\cdot ( 3\vec{k}_2+\vec{q}_2),
\end{equation}

\begin{equation} 
 i L_{c} = \pm {{D+F}\over{2}} {{\sqrt{2}}\over{12 f^3}} {{1}\over{\sqrt{2}}}
 \; 4\;
 \vec{\sigma}\cdot \vec{q}_1, 
\end{equation}
with a plus(minus) sign for the proton(neutron) case,

\begin{equation} 
  i L_{d} =  {{D+F}\over{2}} {{\sqrt{2}}\over{12 f^3}} \; 4 \;
 \vec{\sigma}\cdot\vec{q}_1,
\end{equation}

\begin{equation} 
  i L_{e} =  {{D+F}\over{2}} {{\sqrt{2}}\over{12 f^3}} \; 4 \;
 \vec{\sigma}\cdot\vec{q}_2,
\end{equation}

\begin{equation} 
  i L_{f} = 0
\end{equation}

\newpage

\noindent
{\bf Figure captions:}

\noindent
Fig. 1:

Diagrammatic representation of the $\pi \pi$ scattering matrix contained
in the Lippmann Schwinger coupled channel equations.

\noindent
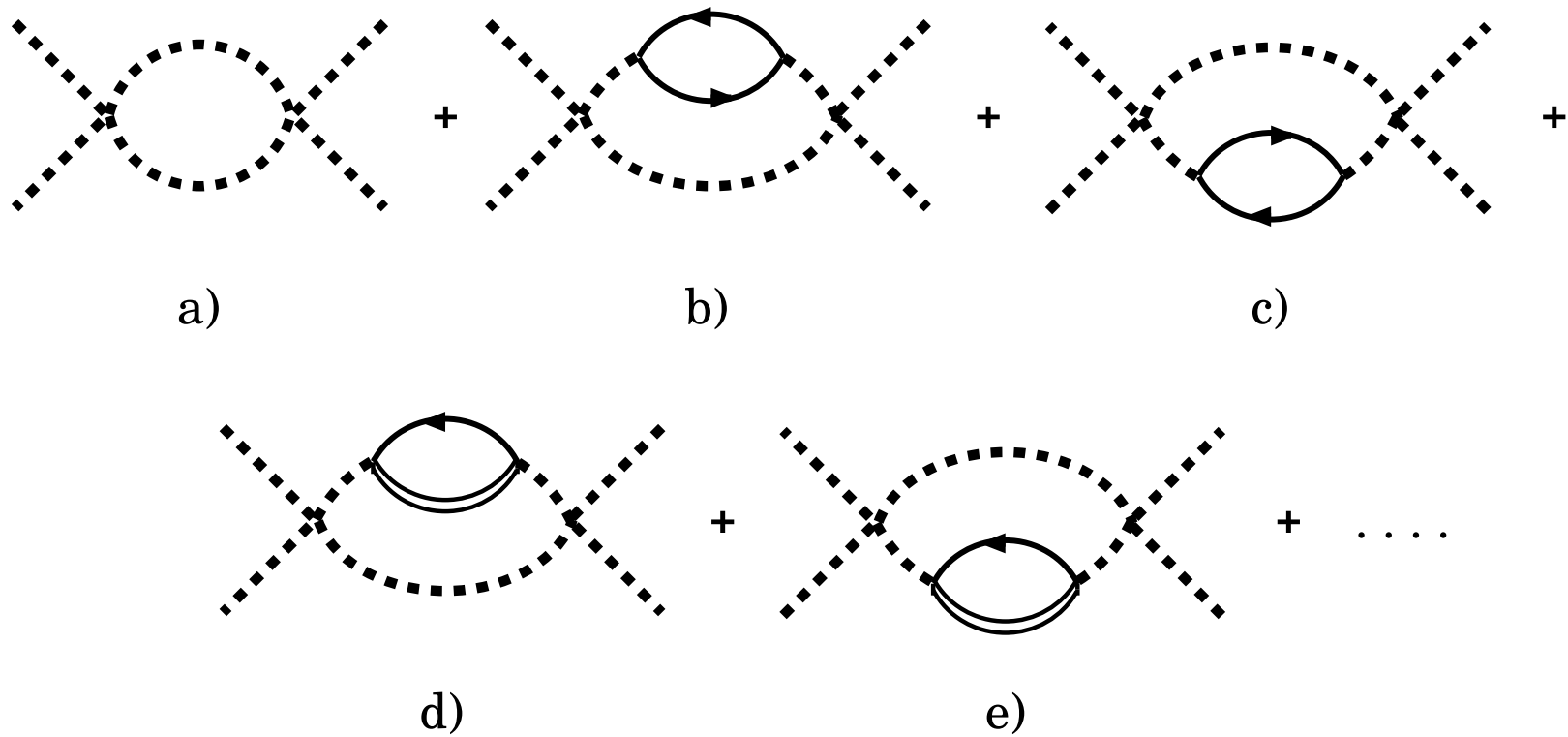
Fig. 2:

Terms appearing in the scattering matrix allowing the pions excite 
$ph$ and $\Delta h$ components.

\noindent
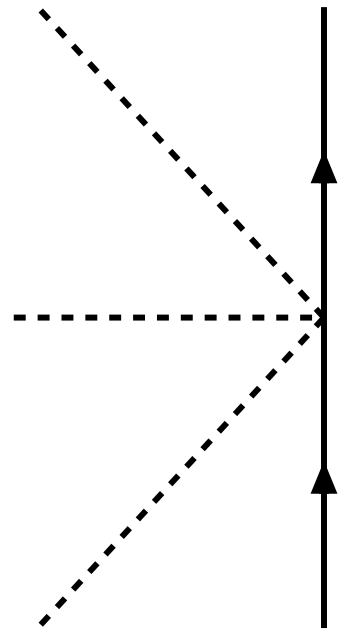
Fig. 3:

Pion pole (a) and contact term (b), appearing in the construction of the
$\pi N \rightarrow \pi \pi N$ amplitude.

\noindent
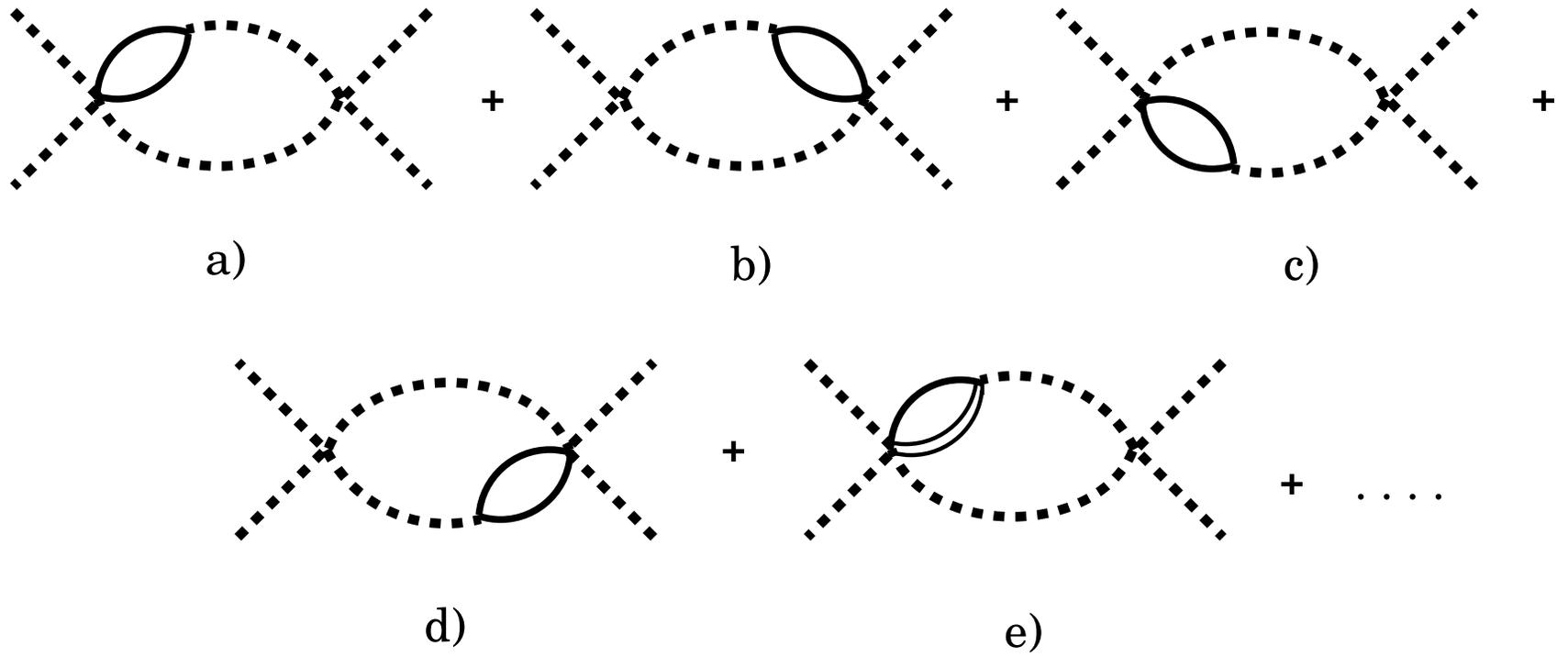
Fig. 4:

Terms of the $\pi \pi$ scattering series in the nuclear medium tied
up to the contact terms of fig. 3.

\noindent
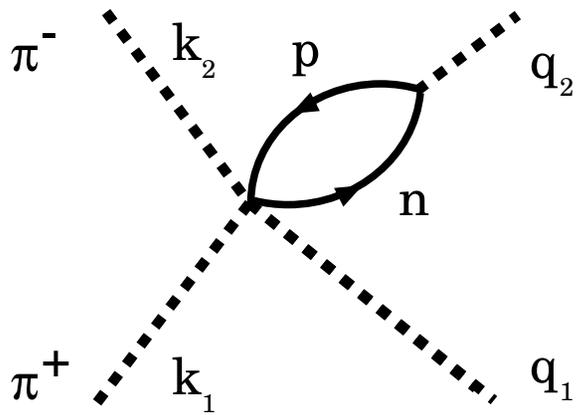
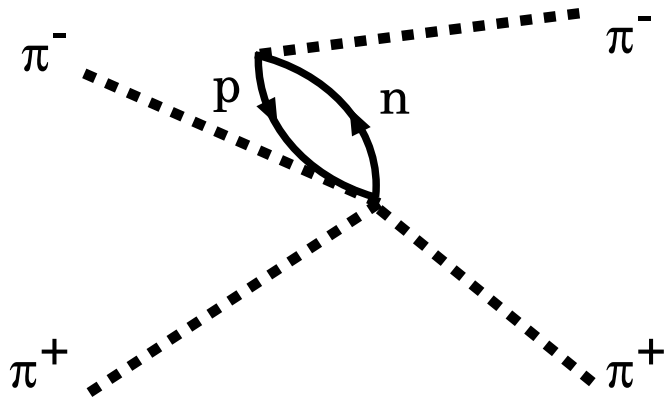
Fig. 5:

Direct and crossed $ph$ excitation terms contained in the 
modified $4 \pi$ vertex, 
accounted for by means of the ordinary Lindhard function.

\noindent
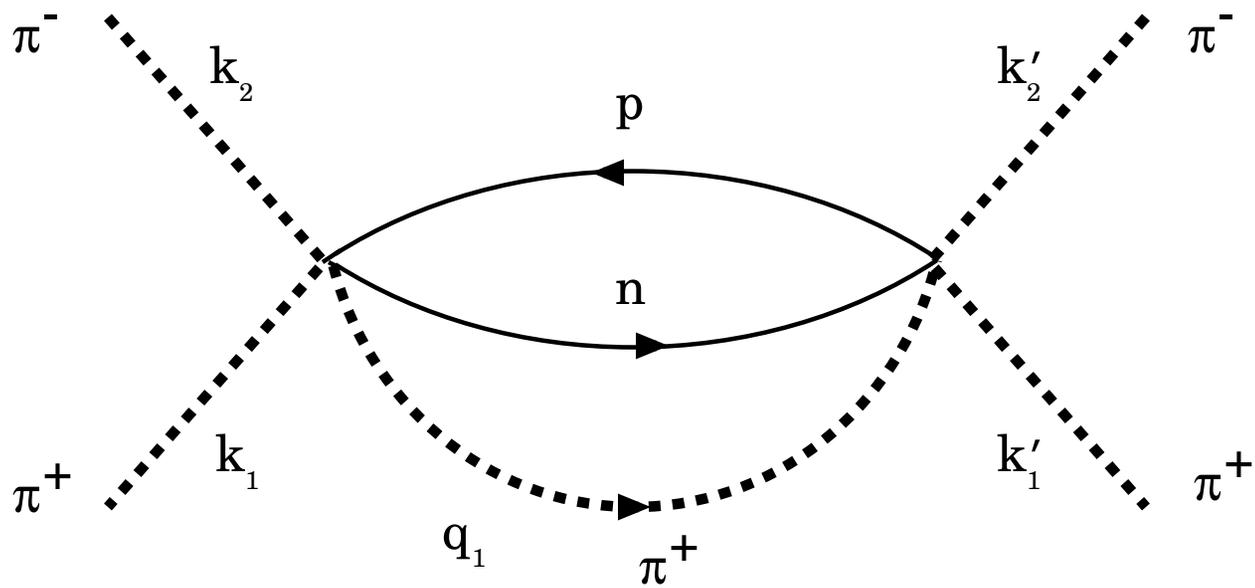
Fig. 6:

Diagram tied to the contact term of fig. 3b, allowing for $ph$ excitation 
and a pion  in the intermediate state.

\noindent
Fig. 7:

Im $T_{22}$  for $\pi \pi \rightarrow \pi \pi $ scattering in $J=T=0$
 $(T_{00}$ in the figure) in the
nuclear medium for different values of $k_F$ versus the CM energy of the 
pion pair. The labels correspond to the values of $k_F$ in MeV.

\noindent
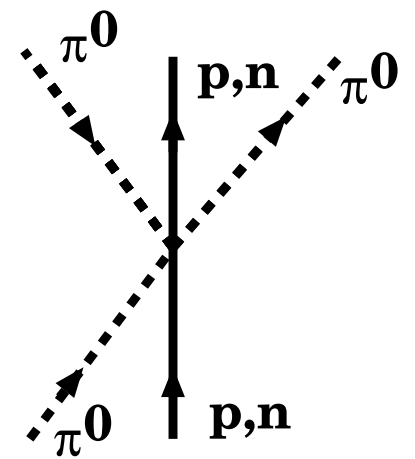
Fig. 8:

Contact terms appearing in the construction of the
$\pi N \rightarrow \pi \pi N$ amplitude.


\newpage

\begin{thebibliography}{99}
\bibitem{1} P. Schuck, W. N\"{o}renberg and G. Chanfray, Z. Phys.
A330 (1988) 119.
\bibitem{2} N. Grion et al., Phys. Rev. Lett. 59 (1987) 1080.
\bibitem{3} F. Bonutti et al., Phys. Rev. Lett. 77 (1996) 603.
\bibitem{4} G. Chanfray, Z. Aouissat, P. Schuck and W.
N\"{o}renberg, Phys. Lett. B256 (1991) 325.
\bibitem{5} J.A. Johnstone and T.S. H. Lee, Phys. Rev. C34 (1986) 243.
\bibitem{6} J. W. Durso, A. D. Jackson and B. J. Verwest,
Nucl. Phys. A345 (1980) 471.
\bibitem{7} Z. Aouissat, G. Chanfray and P. Schuck, Mod. Phys.
Lett. A8 (1993) 1379.
\bibitem{8} Z. Aouissat, R. Rapp, G. Chanfray, P. Schuck and 
J. Wambach, Nucl. Phys. A581 (1995) 471.
\bibitem{9} D. Lohse, J.W. Durso, K. Holinde and J. Speth, Phys. Lett.
B234 (1989) 235; Nucl. Phys. A516 (1990) 513.
\bibitem{10} R. Rapp, J.W. Durso and J. Wambach, Nucl. Phys. A596
(1996) 436.
\bibitem{11} J. Gasser and H. Leutwyler, Nucl. Phys. B250 (1985) 465.
\bibitem{12} A. Dobado, M.J. Herrero and T.N. Truong, Phys. Lett.
B235 (1990) 129; A. Dobado and J.R. Pelaez, Phys. Rev. D47 (1992) 4883;
ibid, Phys. Rev. D56 (1997) 3057.
\bibitem{13} J.A. Oller and E. Oset, Nucl. Phys. A620 (1997) 438.
\bibitem{14} T. N. Truong, Phys. Rev. Lett. 61 (1988) 2526; ibid 67 
(1991) 2260.
\bibitem{14a} J. A. Oller, E. Oset and J. R. Pel\'aez, Phys. Rev. Lett.
80 (1998) 2452.
\bibitem{14b} J. Nieves and E. Ruiz-Arriola, nucl-th/9807035.
\bibitem{15} S. Weinberg, Phys. Rev. Lett. 17 (1966) 616; Phys. Lett.
18 (1967) 188.
\bibitem{16} M.G. Olsson and L. Turner, Phys. Rev. Lett. 20 (1968) 1127.
\bibitem{17} E. Oset, C. Garc\'{\i}a-Recio and J. Nieves, Nucl. Phys.
A584 (1995) 653.
\bibitem{18} U. G. Meissner, E. Oset and A. Pich, Phys. Lett. B353 (1995) 161.
\bibitem{19} E. Oset and M.J. Vicente Vacas, Nucl. Phys. A446 (1985)
584.
\bibitem{19a} O. Jaekel, M. Dillig, C. A. Z. Vasconcelos,
 Nucl. Phys. A541 (1992)673.
584.
\bibitem{20} V. Bernard, N. Kaiser and U.G. Meissner, Nucl. Phys. 
B457 (1995) 147; ibid A619 (1997) 261.
\bibitem{21} T.S. Jensen and A.F. Miranda, Phys. Rev. C55 (1997) 1039.
\bibitem{22} E. Oset and M.J. Vicente-Vacas, Phys. Lett. B386 (1996) 39.
\bibitem{23} U.G. Meissner, Rep. Prog. Phys. 56 (1993) 903; V. Bernard,
N. Kaiser and U.G. Meissner, Int. J. Mod. Phys. E4 (1995) 193.
\bibitem{24} A. Pich, Rep. Prog. Phys. 58 (1995) 563.
\bibitem{25} G. Ecker, Prog. Part. Nucl. Phys. 35 (1995) 1.
\bibitem{26} A. L. Fetter and J.D. Walecka, Quantum Theory of many-particle
systems (McGraw-Hill, New York, 1971).
\bibitem{27} E. Oset, P. Fern\'andez de C\'ordoba, L.L. Salcedo and R.
Brockmann, Phys. Reports 188 (1990) 79.
\bibitem{28a} G. Chanfray and D. Davesne, nucl-th/9806086.
\bibitem{28b} A. Ramos, E. Oset and L.L. Salcedo, Phys. Rev. C50 (1994)
2314.
\bibitem{28c} Z. Aouissat et al., nucl-th/9806069.
\bibitem{29} M. Hermann, B.L. Friman and W. N\"{o}renberg, Nucl. Phys.
A560 (1993) 411.
\bibitem{30} J.V. Steele, H. Yamagishi and I. Zahed, Nucl. Phys.
A615 (1997) 305.


\end{thebibliography}
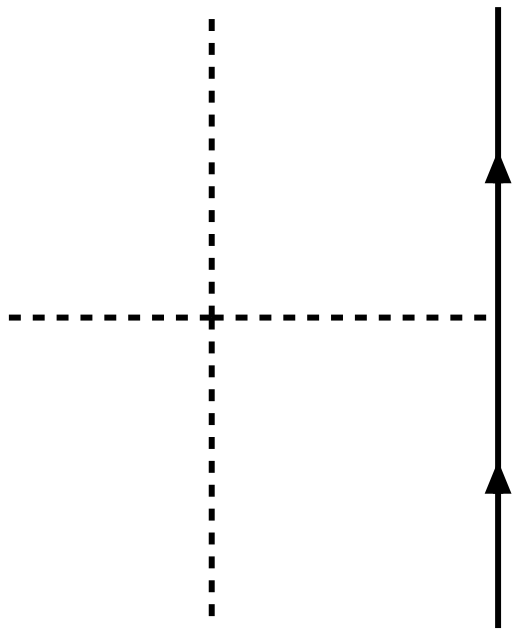
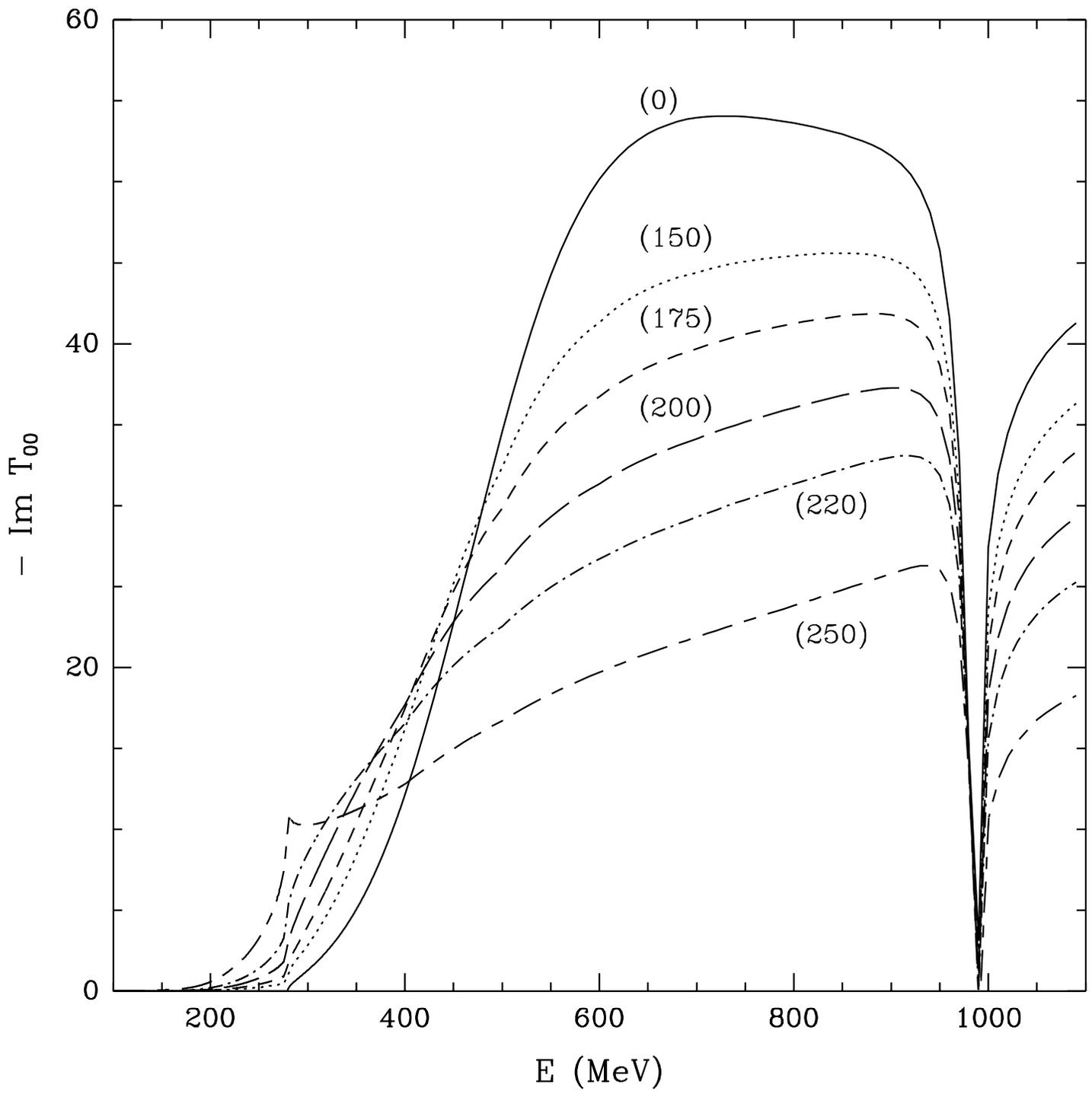
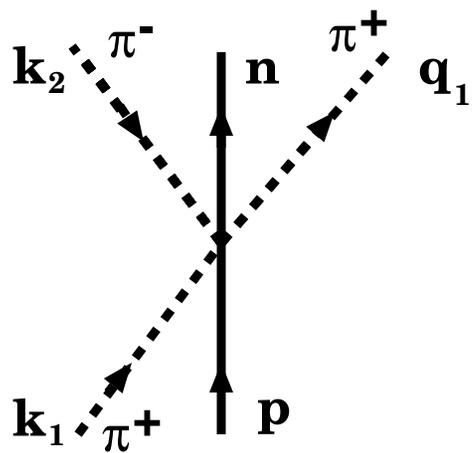
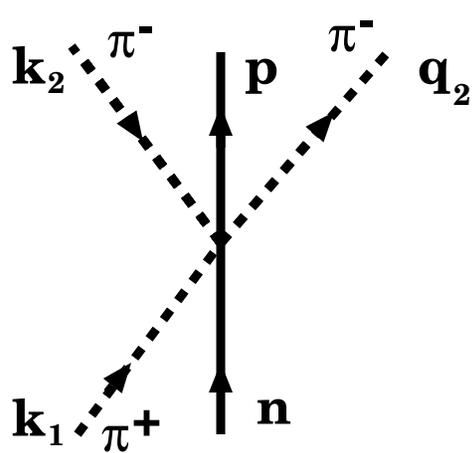
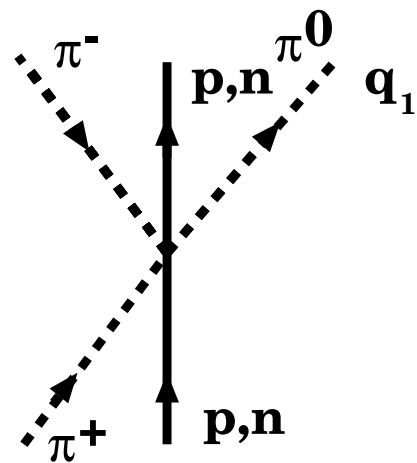
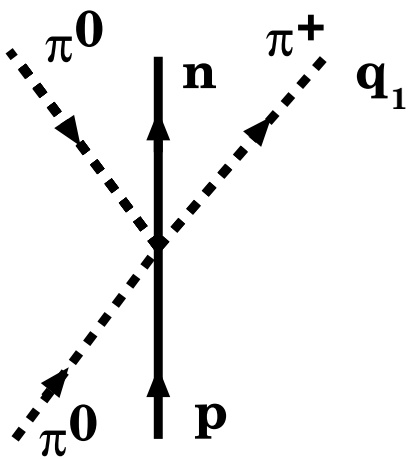
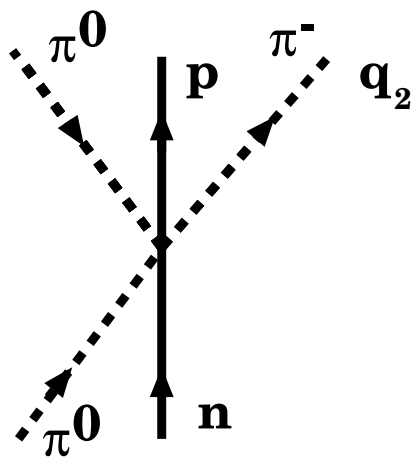
\end{document}